# Interpreting tunneling time in circularly polarized strong-laser ionization

MingHu Yuan[1], PeiPei Xin[1], TianShu Chu[2,3], and HongPing Liu[1*]

[1]*State Key Laboratory of Magnetic Resonances and Atomic and Molecular Physics, Wuhan Institute of Physics and Mathematics, Chinese Academy of Sciences, Wuhan, 430071, China*

[2]*State Key Laboratory of Molecular Reaction Dynamics, Dalian Institute of Chemical Physics, Chinese Academy of Sciences, Dalian, 116023, China.*

[3]*Institute for Computational Sciences and Engineering, Laboratory of New Fiber Materials and Modern Textile, The Growing Base for State Key Laboratory and College of Physics, College of Chemical Science and Engineering, Qingdao University, Qingdao 266071, China*

We propose a method to study the tunneling process by analyzing the time-dependent ionization yield in circularly polarized laser. A numerical calculation shows that for an atom exposed to a long laser pulse, if its initial electronic state wave function is non-spherical symmetric, the delayed phase shift of the ionization rate vs. the laser cycle period in real time in the region close to the peak intensity of the laser pulse can be used to probe the tunneling time. In this region, an obvious delay phase shift is observed, showing the tunneling time is in order of tens of attoseconds. Further study shows the atom has a longer tunneling time in the ionization under a shorter wavelength laser pulse. In our method, a Wigner rotation technique is employed to numerically solve the time-dependent Schrödinger equation of a single-active-electron in a three dimensional spherical coordinate system.

---

* liuhongping@wipm.ac.cn

## I. INTRODUCTION

Strong-field ionization presents a unique combination of quantum and classical physics in atomic and molecular physics. This ionization involves two step: First, the electron is freed from the atom or molecular via tunnel or multi-photon ionization [1], as shown in Fig. 1(a). Then, it propagates classically in the combined ion-laser field. In tunneling which is defined as horizontal ionization, electron tunnels through a potential barrier formed by the Coulomb and laser field without absorbing any photons [2]. In the multi-photon regime defined as vertical ionization, the electron absorbs some photons, thereby gaining enough energy to overcome the potential barrier when $n\omega > I_{ion}$, where $n$ is the photon number, $\omega$ is the laser angular frequency, and $I_{ion}$ is the ionization potential [3]. Unless otherwise stated, atomic units are utilized. The two regimes are usually distinguished by the Keldysh parameter $\gamma = \omega\sqrt{2I_{ion}(\varepsilon^2+1)}/F_0$, where $F_0$ is the peak intensity of the laser electric field, and $\varepsilon$ is the ellipticity [4]. The tunneling regime is characterized by $\gamma \ll 1$, and the multi-photon regime by $\gamma \gg 1$. In the region around $\gamma \approx 1$, however, the mechanism of ionization is unclear yet [5], and the tunneling and multi-photon ionizations can occur simultaneously in this "cross-over" region.

In the ionization process, tunnel is one of the primary manifestations of quantum mechanics departing from classical physics. There are three theoretical assumptions to interpret experimental results of tunneling ionization [6-9]: (A1) First, the highest probability for the electron to tunnel is at the peak of the electric field. (A2) Second, ionization is assumed to be completed once the electron emerges from the barrier. (A3) Third, the photoelectron moves in the combined ion-laser field as a classical particle, and the point of exit and initial distribution is doubtless [1,10]. However, the mechanism of tunneling ionization is unclear. One of the most important topic is the tunneling time. Does the tunneling process consume real time? By far, there is no non-controversial conclusion, experimentally or theoretically.

The attoclock technique had been used for the prior measurement of tunneling time in strong field ionization [7] within a intensity range of $2.3 \times 10^{14}$ to $3.5 \times 10^{14}$ W/cm$^2$, and the experimental results shown that there is no real tunneling delay time. Afterwards, Pfeiffer *et al.* also did not obtain any tunneling delay time using the same technique in helium and argon extended towards higher intensities [9]. At the same time, another experiment measured the tunneling time in the high harmonic generation, but no time could be extracted in this process [11,12]. Recently, Torlina *et al.* reported that no tunneling delays arise in the ionization of single-electron atom, but for the two-electron or multi-electron systems, the interaction of different electrons leads to additional delays [13].

In theory, there are mainly four definitions of the tunneling time named as Larmor time [14,15]; Büttiker-Landauer time [16]; Eisenbud-Wigner time [17] and Pollack-Miller time [18]. The first two definitions depend on the height of the potential and the tunneling times have been called as the resident time; and the other two depend on the incident energy of the particle and the tunneling time have been called as the passage time [3]. On the other hand, tunneling time can also be viewed as average values, rather than deterministic quantities [19,20]. Landsman *et al.* have predicted the time by the probability distribution of tunneling times constructed by using a Feynman Path Integral formulation [21].

As a new powerful tool, circularly polarized laser pulse, has advantages for investigating electron dynamics [22,23]. In circularly polarized laser field, some phenomena are unique, such as angular shifts in photoelectron momentum distributions [24]. Circularly polarized pulse is requisite in the attoclock technique for measurement of tunneling time, which is also employed in the present work. In this paper, we theoretically investigate the tunneling time by analyzing the time-dependent ionization yield in circularly polarized laser field. Taking argon atom as an example, its electron density of initial ground state looks like pea and the maximum is along $z$ axis as shown in Fig. 1(b). In a laser cycle period, there are two maximum ionization yields corresponding to the two electron density maximum. The ionization yield should get the largest value when the electric field points in the $z$ axis where the electron density is maximum [4,25]. However, in some special laser fields, the maximum ionization yield occurs with an observable delay time, which serves a probe to detect the time consumed in the tunneling ionization process. In our work, employing the spectral method in length gauge, we solve the three dimensional time-dependent Schrödinger equation (TDSE) [26] and study the dependence of ionization yield on the applied pulse. In the long pulse case, we find an obvious delay in order of tens of attoseconds which is caused by the tunneling time.

## II. THEORETICAL METHOD

The Wigner rotation technique is introduced for solving the TDSE in the theoretical method which has been described in detail in Ref. [27] and we just give a brief overview. Taking the spherical symmetry of atomic system into account, we choose the spherical coordinate $\mathbf{r}=(r,\theta,\phi)$ in the calculation and adopt the single-active-electron model to describe the dynamics of atom in strong laser field. Then, the time-dependent wave function can be expanded as [28],

$$\psi(\mathbf{r},t) = \sum_{l=0}^{l_{max}} \sum_{m=-l}^{l} \frac{1}{r} \chi_{l,m}(r,t) Y_{l,m}(\theta,\phi). \tag{1}$$

Here, the reduced radial wave function $\chi_{l,m}(r,t)$ is represented on the basis of Sine-DVR (Sine basis functions are used to define the discrete variable representation) [29], and $Y_{l,m}(\theta,\phi)$ is the spherical harmonic. Based on this representation, we can benefit from angular momentum theory when dealing with the angular degrees of freedom.

The TDSE of single-active-electron can be written as,

$$i\frac{\partial}{\partial t}\psi(\mathbf{r},t) = \left(-\frac{\nabla^2}{2} + V^{(a)}(r) + V^{(F)}(\mathbf{r},t)\right)\psi(\mathbf{r},t), \tag{2}$$

where $V^{(a)}(r)$ represents the spherically symmetric three-dimensional potential of atomic system and $V^{(F)}(\mathbf{r},t) = \mathbf{r} \cdot \mathbf{F}(t)$ is the laser-atom interaction under dipole approximation. In order to solve the three dimensional TDSE efficiently, we take advantage of the Wigner rotation technique which has been introduced in Ref. [30,31]

in detail. As the circularly polarized light can be deemed to the rotation of the linearly polarized light, we propagate the time-dependent wave function in linearly polarized case and rotate the wave function by Wigner rotation matrix in each step of the process. Thus, the linearly polarized laser is equivalent to circularly polarized laser to the revolving atomic system.

The second-order split-operator scheme is employed to propagate the wave function fast and efficiently [32],

$$\chi_{l,m}(t+\delta t) = e^{-\frac{\delta t}{2}T_r} e^{-i\frac{\delta t}{2}T_l} e^{-i\frac{\delta t}{2}V^{(a)}(r)} e^{-i\delta t V^{(F)}} \\ \times e^{-i\frac{\delta t}{2}V^{(a)}(r)} e^{-i\frac{\delta t}{2}T_l} e^{-i\frac{\delta t}{2}T_r} \chi_{l,m}(t) + O(\delta t^3) \quad . \quad (3)$$

Here, $T_r = -(1/2)\partial^2/\partial r^2$, $T_l = l(l+1)/2r^2$ and $V^{(a)}(r)$ are radial kinetic operator, centrifugal operator and potential operator, respectively, which are related to the interaction of electron with nuclei. And $V^{(F)}$ is the electron-field interaction operator.

It is convenient to treat the interaction of electron with the nuclei and the field separately at each step of the time propagation. So the wave function should be rotated only when we treat the interaction of electron and field. In the time propagation, every step can be split into three sub-steps. First, we represent the wave function in the atomic frame and calculate the action of electron-nuclei interaction operators (the last three terms in Eq. (3)). Second, we revolve the updated wave function by the Wigner rotation matrix $\mathbf{D}(\beta)$, and then apply the obtained wave function to the electron-field interaction operator (the middle term in Eq. (3)). The element of Wigner rotation matrix is represented as,

$$D_{m,m'}^{l}(\beta) = \sqrt{\left[(l+m')!(l-m')!(l+m)!(l-m)!\right]} \\ \times \sum_{v} \left\{ \frac{(-1)^{v}}{(l-m-v)!(l+m'-v)!(l+m-m')!v!} \right. \\ \left. \left(\cos\left(\beta/2\right)\right)^{2l+m'-m-2v} \left(-\sin\left(\beta/2\right)\right)^{m-m'+2v} \right\} \quad, \quad (4)$$

where $v \in \left[\max(0, m'-m), \min(l-m, l+m')\right]$ and $\beta$ is the rotation angular. Finally, we transform the electronic wave function back by inverse rotation again to act on the electron-nuclei interaction operators (the first three terms in Eq. (3)). Then a complete step of the time propagation is achieved.

It should be noted that the Wigner rotation matrix is block diagonal with respect to $l$. Thus, in the rotation process, there is no mix between different $l$ states. At the same time, when we treat the interaction of electron with the nuclei, we do not mix different $m$ states discussed previously [28]. Therefore, in the whole propagation, we avoid the mix of different $l$ and $m$ states at the same time, which means that we reduce the three-dimensional problem to a number of two-dimensional problems.

The ionization yield is defined as the follows. The projection of time-dependent wave function $\psi(\mathbf{r},t)$ onto the bound eigenstates $\psi_j(r,\theta,\phi) = (1/r)\chi_{j,l=1}(r)Y_{l=1,0}(\theta,\phi)$ corresponds to the electrons remaining at the bound states at evolution time $t$, where the eigenstates can be obtained by solving the time-independent Schrödinger equation by diagonalizing the field-free Hamiltonian. Therefore, the ionization yield can be written as follows:

$$P_{ioni}(t) = 1 - \sum_j \left| \int_{r_{min}}^{r_{max}} \int_0^{\pi} \int_0^{2\pi} \psi_j^*(r,\theta,\phi) \psi(r,\theta,\phi,t) r^2 \sin\theta dr d\theta d\phi \right|^2$$

$$= 1 - \sum_j \left| \sum_{l=0}^{l_{max}} \sum_{m=-l}^{l} \int_{r_{min}}^{r_{max}} \chi_{j,l=1}^*(r) \chi_{l,m}(r,t) dr \right|^2 , \quad (5)$$

where the ionization yield is time-dependent and can be used to evaluate the dynamics of electron with the laser pulse in real time.

### III. NUMERICAL RESULTS AND DISCUSSION

There are two empirical formulas for the three dimensional atomic potential of argon atom [33,34], and both of them are accurate and used widely [35-37]. In our calculation, the atomic potential is taken from Ref. [34]. The radial spatial interval is about 3.8 a.u. in the range of [0.0, 400 a.u.]. We choose the ground electronic state ($3p_z$) of argon atom as the initial state, and the wave function is shown in Fig. 1(b). We do not consider the influence of different magnetic quantum states, and the initial state is chosen to be the one of $m=0$. The expansion in spherical harmonics is truncated at $l \leq l_{max}$ with $l_{max} = 55$, which is satisfied for convergence in our calculations. The propagation time step is 0.02 a.u. The circularly polarized laser pulse is assumed to oscillate in the $xz$ plane in Cartesian coordinate system, that is, $\theta = 0$ or $\pi$ and $\phi = 0$ or $\pi$ in spherical coordinate system, and the electric field is decomposed into two vectors: $\mathbf{F}_x(t) = F_0 f(t) \cos(\omega t + \varphi) \mathbf{e}_x$ that parallels the $x$ axis; $\mathbf{F}_z(t) = F_0 f(t) \sin(\omega t + \varphi) \mathbf{e}_z$ that parallels the $z$ axis, with $\varphi$ the carrier-envelope phase (CEP), $f(t) = \sin^2(\omega t / 2N) / \sqrt{2}$ the envelope and $N$ the number of the optical cycles.

In order to study the tunneling process, we have to choose an appropriate laser pulse to make sure the ionization dynamics occurs in the tunneling regime. A typical photoelectron energy spectrum (PES) of argon atom exposed in circularly polarized laser pulse with wavelength of 800 nm is shown in Fig. 2. The intensities are $1.0 \times 10^{14}$ W/cm² and $1.8 \times 10^{14}$ W/cm² corresponding to the ponderomotive energy of $U_p = 6.0$ eV and $U_p = 10.8 \, \text{eV}$, respectively. Here, $U_p = F_0^2/(2\omega)^2$. There is a main peak in both of the two PES curves, and the position of the peak is at around the ponderomotive energy. The feature implies the atom is ionized by laser field in tunneling regime and all our calculations thereafter are performed in this condition.

In the tunneling regime, we can trace the ionization yield in real time as shown in Fig. 3(a, c), where the corresponding applied circularly polarized pulses with different CEPs are displayed as well in Figs. 3(b, d). Here, the laser intensity is fixed at $1.0 \times 10^{14}$ W/cm², the wavelength equals to 800 nm corresponding to the angular frequency $\omega = 0.057$ a.u., and the laser pulse duration is 3 optical cycles. The time-dependent ionization yield is dominated by Coulomb field and laser electric field at the same time. According to the theory of tunneling and the form of initial electron density in a circularly polarized field [25], the ionization yield as a function of time should be oscillating, and interval of peaks is about half a cycle. For the case of $\varphi = 0$, there is a peak in the ionization yield curve [peak a in Fig. 3(a)] at 1.5 cycle when the electric field points in the direction of $\theta = 180°$ where is a maximum of electron density of the $3p_z$ state as shown in Fig. 3(b) [the white solid arrow indicates the electric field at this moment]. However for the case of $\varphi = \pi/2$, there is a valley in the curve [valley d in Fig. 3(c)] at 1.5 cycle when the electric field points in the direction of the minimum of electron density, and the peaks [peaks b and c in Fig. 3(c)] appear in the vicinity of 1.25 and 1.75 cycle. Thus, the time-dependent ionization yield is influenced mainly by the initial electronic state density, and the maximum ionization probability is obtained near the moment when the electric field points in the direction of $\theta = 0°$ or $180°$ where the electron density arrives at the densest.

We can also draw the same feature on the time-dependent ionization yields at other CEPs of $\varphi = 0$, $\pi/6$, $\pi/3$ and $\pi/2$, as shown in Fig. 4. It is obvious that the peaks in the curves move left meaning that the largest ionization yield appears earlier as the CEP increases. And the interval of displacement of peaks approximate the time corresponding to the phase difference $\Delta d = \Delta\varphi/2\pi$, where $\Delta d$ (cycle) is the time of displacement of peaks and $\Delta\varphi$ is the phase difference. Thus, we can investigate the ionization dynamics by ascertaining the accurate position of the maximum ionization probability.

This character is valid at other laser intensities as well. Figure 5 exhibits the time-dependent ionization yields in 3 cycles circularly polarized laser pulse with intensities of $0.5 \times 10^{14}$ W/cm$^2$ and $1.0 \times 10^{14}$ W/cm$^2$, respectively. The wavelength is 800 nm, the same as previous calculations shown in Fig.3-4. All the numerical results show that a local maximum of ionization yield occurs when the laser electric field points in the direction where the atomic electron density is maximum (peaks a-c in

Fig.3 and peaks a-c in Fig.5). However, carefully observing the ionization yields shown in Fig.3 and Fig. 5, we notice that there are slight time offsets from the regular time for some peaks of ionization yield (peaks b, c in Fig.3 and peaks a, c in Fig.5), where the laser pulse intensity is increasing or decreasing with time. And only the middle peaks (peak a in Fig.3 and peak b in Fig.5) occur punctually. Another feature is that the time deviation for peaks b, c in Fig.3 and peaks a, c in Fig.5 are symmetric around the middle time, that is, their time phase shifts are equal but own opposite signs. This implies that the time shift is due to the effect of ultrashort laser pulse, that is, the higher ionization probability appears at the moment when the laser is stronger and the intensity varies violently in ultrashort laser pulse.

If applying a long laser pulse to ionize the atom, we can reduce or remove the effect of ultrashort laser pulse and use it as a probe to monitor the ionization dynamics more accurately in time, for example, the tunneling ionization time. We calculate the time-dependent ionization yield in 9 cycles circularly polarized laser pulse with wavelength 800 nm and intensity $1.8 \times 10^{14}$ W/cm$^2$ in Fig. 6(a). The effects of ultrashort laser pulse can be neglected for the middle peaks around the center time of pulse (the position at 4.5 cycle), as the electric field varies slowly. And if the ionization is affected only by the effects of ultrashort laser pulse, the middle peak will appears at the middle of the pulse. However, we find an obvious time offset as magnified in the inset. The offset corresponds to the tunneling time. Thus, the tunneling time can be obtained using the relationship $\Delta t = \Delta d \cdot C_t$, where the $C_t$ is the pulse cycle time. In this case, we get the tunneling time $\Delta t$ equal to 44 as. As all know, there are two peaks for the electron wave function of $3p_z$ state as shown in Fig. 1(b). Thus, two peaks and two valleys emerge in the time-dependent ionization yield curves for every pulse cycle, and there should be shift for every peak. In the following calculation, we extract all the offsets for the peaks in the middle region of the laser pulse used in Fig. 6(a). The shifts, which mean the delay time when the electric field points to the peak of wave function, are displayed in Fig. 6(b). It is clear that the offsets for the first peaks are positive and negative for the last peaks. Thus it can be seen that the offsets are affected by the effect of ultrashort laser pulse at the both ends of the laser pulse. Therefore, we can check the middle peak (the 6[th] peak in this calculation) to investigate the tunneling time and we find there are obvious delay times. This feature implies that the tunneling time is real.

To remove the effects of ultrashort laser pulse completely, we have a look at the ionization process under a flat-top laser pulse as shown in Fig. 7. The ionization yields in circularly polarized laser field with wavelength of 800 nm and 600 nm are both calculate and displayed in Fig. 7(a), where the laser intensity is $1.8 \times 10^{14}$ W/cm$^2$. In this calculation, we choose 9 cycles flat-top laser pulse with turn-on and turn-off of sine-squared form. And the delay time for peaks in the both curves in flat-top region (the shadow region in Fig. 7(a)) of laser pulses are extracted, as shown in Fig. 7(b), where the delay times are almost constants and it further proves the long pulse laser can be used to study the tunneling process. The tunneling time of 44 as obtained from

Fig. 7(b) at 800 nm has the same value as by a long pulse in Fig. 6. At the same time, we can also find that the delay time is much longer in 600 nm laser pulse (100 as) than that in 800 nm laser pulse, which means that the tunneling process needs more time in short wavelength laser field. We conjecture the reason for this feature is that the tunneling time is influenced by the non-adiabatic effects, on account of that photoelectron will absorb energy in non-adiabatic ionization process [38]. And the non-adiabatic manifestations get stronger with shorter wavelength pulse [39,40]. In the next work, we will investigate the relationship between tunneling time and non-adiabatic effects.

## IV. CONCLUSION

In summary, we propose a method to study the tunneling time by analyzing the time-dependent ionization yield in circularly polarized laser, and apply it to the ionization of argon in strong laser fields. An obvious tunneling time of tens of attoseconds is determined in our condition. Our calculation shows that the offsets of the peaks of the ionization yield in the ultrashort laser pulse, as well as that at both ends of the long laser pulse, are affected by the ultrashort pulse effect, while the offsets of middle peaks in long laser pulse are not. It provides us a way to investigate the tunneling time in the ionization by long pulses. In the end, we find the tunneling process need more time in short wavelength laser field. In the calculation, Wigner rotation technique is used to solve the TDSE of a single-active-electron in a three dimensional spherical coordinate in length regime.

## ACKNOWLEDGMENTS

We gratefully acknowledge L. B. Madsen for the help on the theoretical method. The work was supported by the China Postdoctoral Science Foundation (Grant No. 2015M572229), the National Key Basic Research Program of China (Grant No. 2013CB922003), NSFC (Grants No. 11174329, No.91121005, No. 91421305 and No. 11504412) and Shandong Provincial Natural Science Foundation, China (ZR2014AM025).

**Figures**

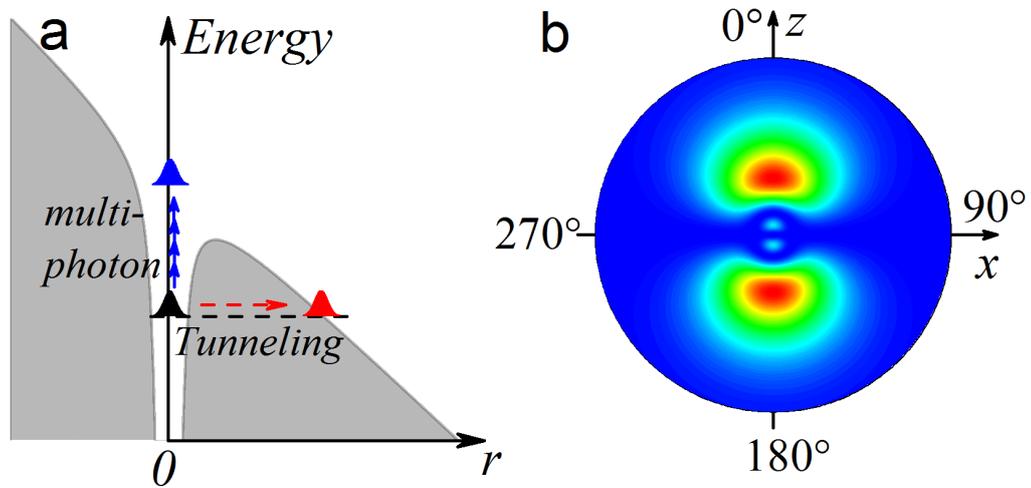

Fig. 1. (Color online) (a) Strong field ionization. The electron can escape the atom either by tunneling (horizontal channel) or multi-photon (vertical channel) ionization. The potential form is created by the Coulomb field and the laser field. (b) The ground state electron density of argon atom ($3p_z$). The wave function is chosen as the initial state in the time-dependent propagation.

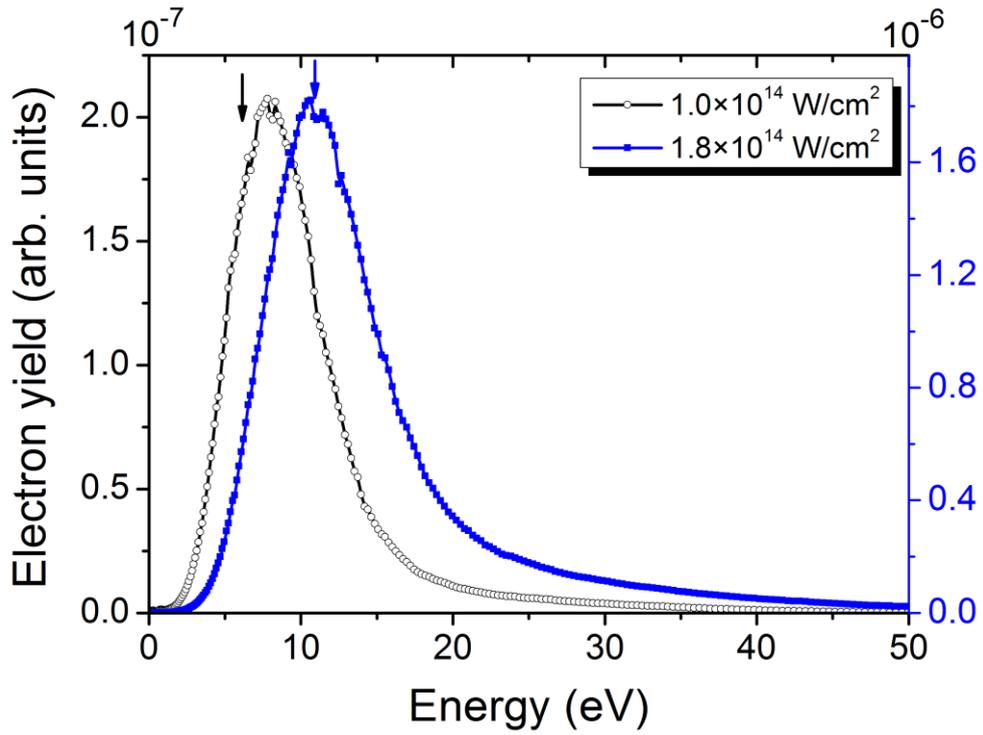

Fig. 2. (Color online) PES of argon atom exposed in 800 nm ($\omega = 0.057$ a.u.) circularly polarized laser pulse containing 3 optical cycles. The intensities are $1.0 \times 10^{14}$ W/cm² and $1.8 \times 10^{14}$ W/cm² corresponding the Keldysh parameters of $\gamma = 1.6$ and $\gamma = 1.2$, respectively. The arrows indicate the ponderomotive energy of $U_p = 6.0$ eV and $U_p = 10.8$ eV.

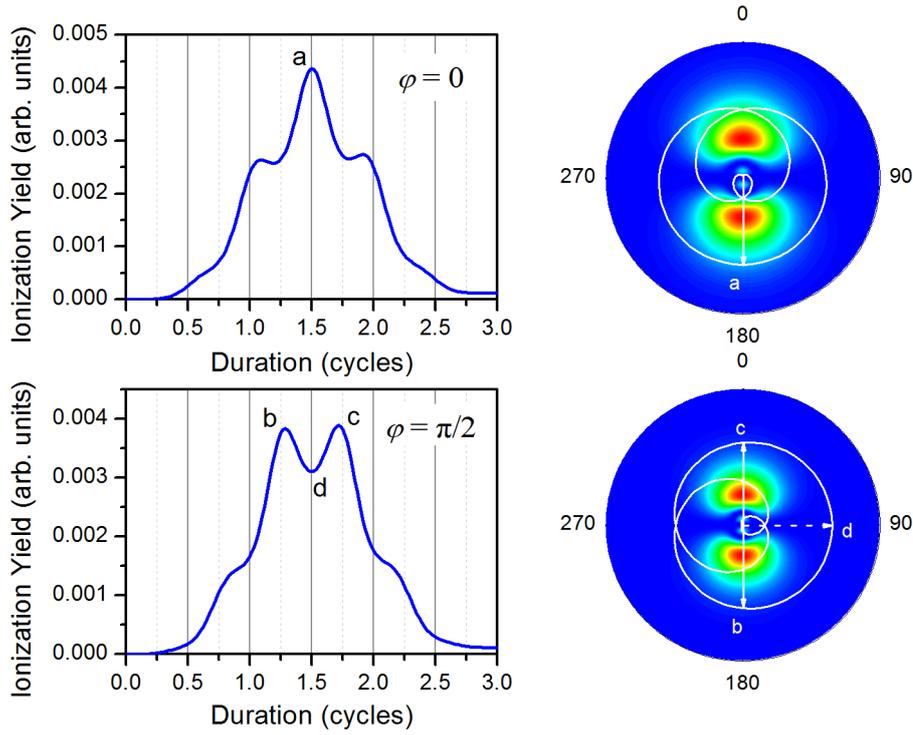

Fig. 3. (Color online) Time-dependent ionization yield (a, c) and the corresponding circularly polarized pules with different CEPs (b, d). The laser pulses contain 3 optical cycles with CEPs of $\varphi = 0$ and $\varphi = \pi/2$, respectively. The laser intensity is $1.0 \times 10^{14}$ W/cm$^2$ and the wavelength is 800 nm. Note that the moment that the maximum of ionization yield occurs deviates slightly from the time that the laser electric field points in the direction where the atomic electron density is maximum for peaks b and c.

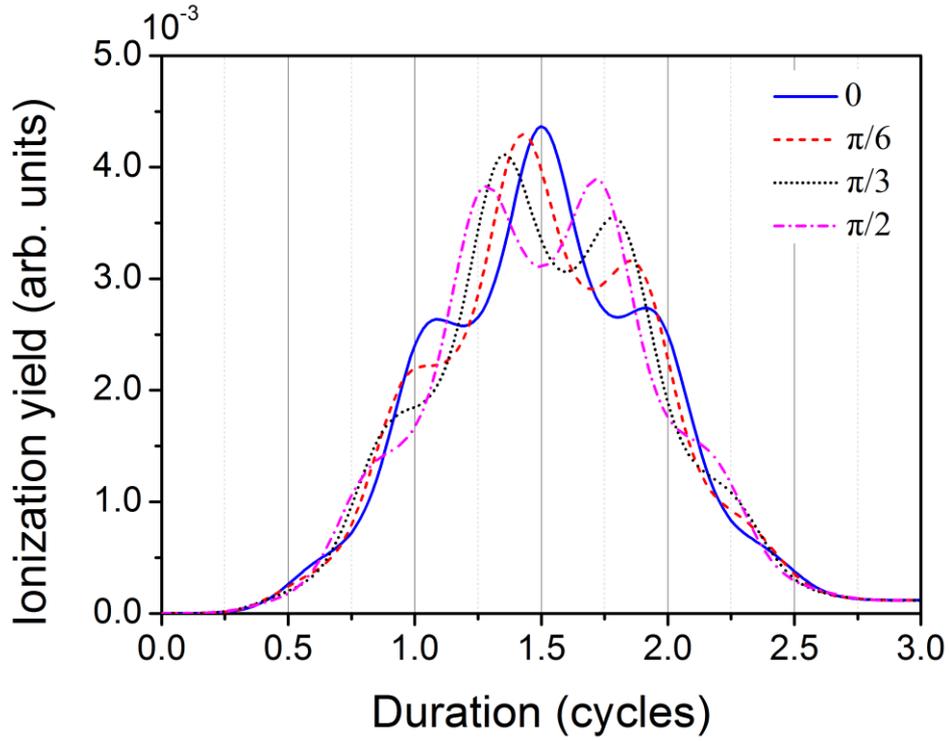

Fig. 4. (Color online) Time-dependent ionization yields of argon atom exposed in circularly polarized fields with CEPs of $\varphi=0$, $\varphi=\pi/6$, $\varphi=\pi/3$ and $\varphi=\pi/2$. The other laser parameters are the same as those in Fig. 3.

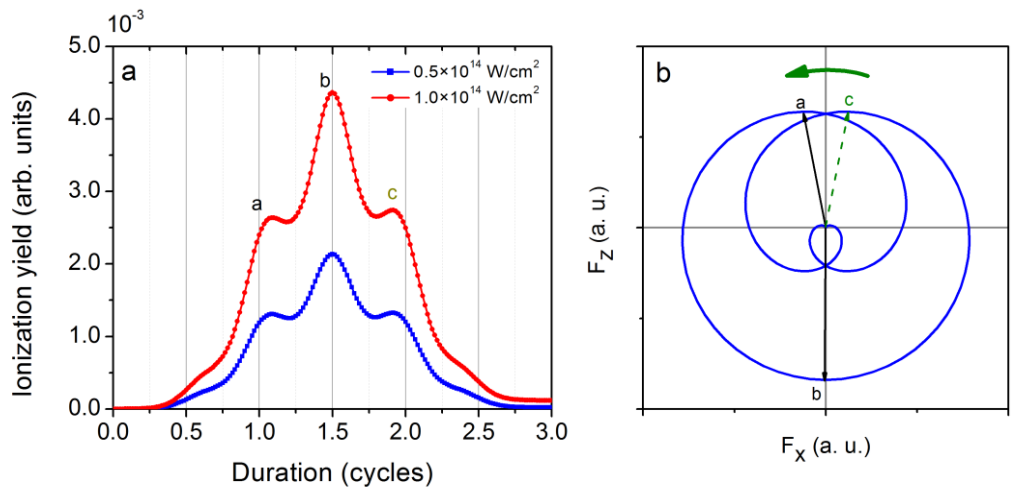

Fig. 5. (Color online) (a) Time-dependent ionization yields in 3 cycles circularly polarized laser pulses with different intensities of $0.5 \times 10^{14}$ W/cm$^2$, $1.0 \times 10^{14}$ W/cm$^2$ and $1.8 \times 10^{14}$ W/cm$^2$, respectively. The wavelength is 800 nm and CEP is $\varphi = 0$. (b) The form of electric field. Note that the moment that the ionization yield peaks a and c occurs deviates from the time that the laser electric field points in the direction where the atomic electron density is maximum.

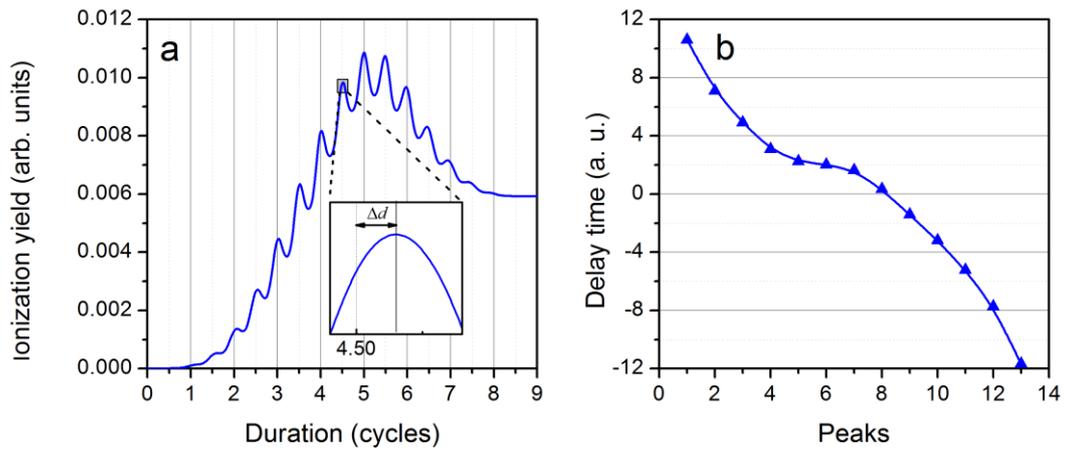

Fig. 6. (Color online) (a) Time-dependent ionization yield in 9 optical cycles circularly polarized laser pulse with wavelength of 800 nm and intensity of $1.8 \times 10^{14}$ W/cm$^2$. (b) The shifts of peaks in time-dependent ionization yield curve from the time when the electric field points to the maximum of wave function.

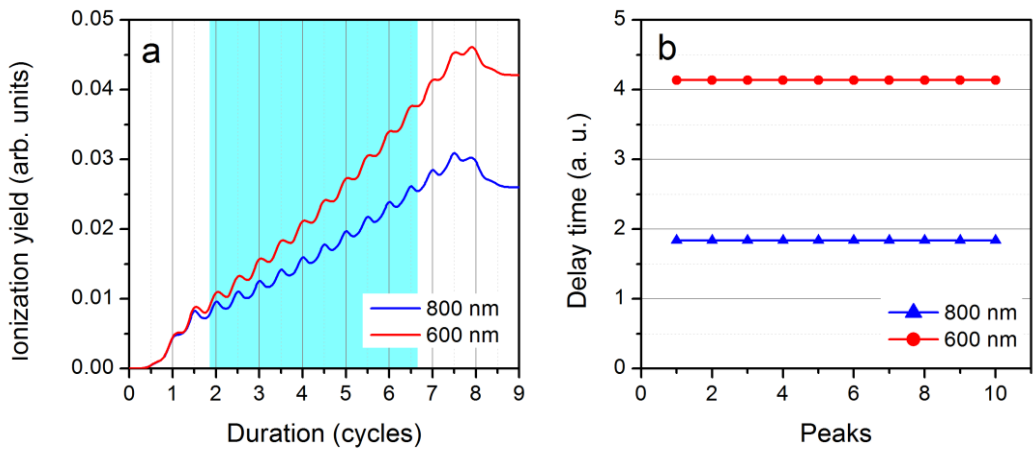

Fig. 7. (Color online) (a) Time-dependent ionization yield in 9-cycle flat-top circularly polarized pulse with different wavelengths. The same laser intensity is $1.8 \times 10^{14}$ W/cm$^2$. And (b) the comparison of the delay times for peaks in flat-top region of the circularly polarized laser pulses.